\documentclass[
    aps,prl,twocolumn,groupedaddress,
    superscriptaddress,amsfonts,
    amssymb,amsmath,citeautoscript,
    longbibliography,letter,10pt
    ]{revtex4-2}

\usepackage[utf8]{inputenc}
\usepackage[english]{babel}

\usepackage{microtype} 
\usepackage{xspace} 

\usepackage{txfonts}  
\usepackage{txfontsb} 
\usepackage{scalefnt}

\usepackage{bm} 

\usepackage{xcolor}
\usepackage[]{graphicx}

\usepackage[]{booktabs}
\usepackage{array}
\usepackage{layouts}
\usepackage{multirow}

\usepackage{enumerate}
\usepackage[inline]{enumitem}

\usepackage{hyperref}
\hypersetup{colorlinks,
	linkcolor={blue!75!black!80!yellow},
	citecolor={blue!75!black!80!yellow},
	urlcolor={blue!75!black!80!yellow}
}

\usepackage[capitalize,nameinlink]{cleveref}

\crefname{subequations}{Eqs.}{Eqs.} 
\Crefname{subequations}{Eqs.}{Eqs.}
\crefformat{subequations}{#2Eqs.~(#1)#3}
\Crefformat{subequations}{#2Eqs.~(#1)#3}
\crefname{page}{p.}{p.} 

\usepackage{placeins}

\usepackage{siunitx}
\DeclareSIUnit[number-unit-product = ]\percent{\char`\%} 

\usepackage[centering,hmargin=16mm,tmargin=30mm,bmargin=26mm]{geometry}

\renewcommand{\paragraph}[1]{\vskip 1ex\noindent\textbf{#1.}~}

\usepackage[eulergreek]{sansmath}
\makeatletter
\renewcommand\@make@capt@title[2]{%
    \@ifx@empty\float@link{\@firstofone}{\expandafter\href\expandafter{\float@link}}%
    \sffamily\textbf{#1\@caption@fignum@sep}#2 
}%

\makeatother
\usepackage{svg}
\usepackage{placeins}
\newcommand{\rv}{\mathbf{r}}

\newcommand{\dd}{\mathrm{d}}

\newcommand{\eg}{e.g.,\@\xspace}

\newcommand{\appropto}{\mathrel{\vcenter{
			\offinterlineskip\halign{\hfil$##$\cr
				\propto\cr\noalign{\kern.2pt}\sim\cr\noalign{\kern-2.5pt}}}}}

\usepackage{gensymb}
\renewcommand{\degree}{\ifmmode ^\circ \else \textdegree \fi}

\begin{document}
\scalefont{1.1}
\title{Steerable Radiation Forces with Frequency-Detuned Acoustic Metasurfaces}


\newcommand{\umnMechE}{Department of Mechanical Engineering, University of Minnesota, Minneapolis, MN 55455, USA}
\newcommand{\umnPhys}{School of Physics and Astronomy, University of Minnesota, Minneapolis, MN 55455, USA}

\author{Sam Keller*}
\author{Matthew Stein*}
\affiliation{\umnMechE}
\author{Ognjen Ilic$^{\dag}$}
\affiliation{\umnMechE}



\begin{abstract}
We demonstrate that acoustic waves can induce controlled translation and rotation of macroscopic objects through small, but deliberate, detuning of the driving wave frequency.
When an object is patterned with a suitably designed acoustic metasurface, small changes in the incident frequency $\omega \pm \delta \omega$ are converted into directional radiation forces and torques, enabling steerable motion even for objects much larger than the acoustic wavelength.
We present the concept of a force-optimal metasurface topology and show that it enables fully reversible forces in real time: the object is moved in one direction for positively detuned incident frequency $\omega+\delta \omega$ and in the opposite direction for negatively detuned frequency $\omega-\delta \omega$, where $\omega=22.5 \textrm{ kHz}$ and $\delta \omega =2.5 \textrm{ kHz}$ for a proof of concept at inaudible frequencies.
This mechanism is demonstrated experimentally at ultrasonic frequencies with 3D-printed metasurfaces. 
The proposed concept is scalable across frequencies and materials, offering a building block for realizing complex, remote-controlled, dynamical behaviors that can be programmed by reconfiguring material surface patterns.
\end{abstract}

\maketitle

\section{Introduction}

The ability to actuate objects remotely---without tethers, wiring, or onboard power---represents a longstanding goal across robotics, active matter systems, and programmable and responsive materials and structures.
Traditionally, actuation has been achieved by embedding actuators and power supplies directly within the moving object, or by relying on physical connections such as mechanical linkages and wires. These approaches, however, inevitably add complexity, limit scalability, and constrain operation in environments where contact is undesirable or impractical.
By contrast, waves, such as acoustic and optical waves, offer a fundamentally different paradigm for actuation.
They deliver forces and torques contactlessly and without requiring onboard energy source.
For small objects, such as particles, this principle underpins acoustic traps and tweezers \cite{Brandt-Brandt:Nature01-n, Baresch-Marchiano:PhysRev16-c, Marzo-Drinkwater:RevSci17-t, Andrade-Adamowski:ApplPhys20-b}.
These and related techniques have been exploited for particle control applications 
\cite{Anhauser-Brasselet:PhysRev12-y, Foresti-Poulikakos:PhysRev14-e, Melde-Fischer:Nature16-l, Baresch-Marchiano:PhysRev18-n, Baresch-Garbin:ProcNatl20-e, Li-Cummer:SciAdv21-m}%
, including sorting and imaging \cite{Yang-Dholakia:NatCommun19-d}%
, haptics \cite{Hoshi-Shinoda:IEEETrans10-j}%
, display technology \cite{Hirayama-Subramanian:Nature19-c, Fushimi-Hill:ApplPhys19-r}%
, and contactless assembly \cite{Vandaele-Delchambre:PrecisEng05-f, Lim-Jaeger:NatPhys19-l}, 
where ultrasonic waves levitate particles and objects small enough to fit between the nodes of the wave.
However, these mechanisms necessitate particles to be small, typically smaller than $\lambda/2$, where $\lambda$ is the wavelength of the incident wave excitation, and extending such wave-based manipulation to larger objects remains a central challenge.
In the regime where object's dimensions exceed the wavelength, conventional approaches fail because such objects cannot be confined between wave nodes, and radiation forces are strongly constrained by the object geometry and curvature. 

\begin{figure*}[ht]
     \includegraphics[width=\textwidth]{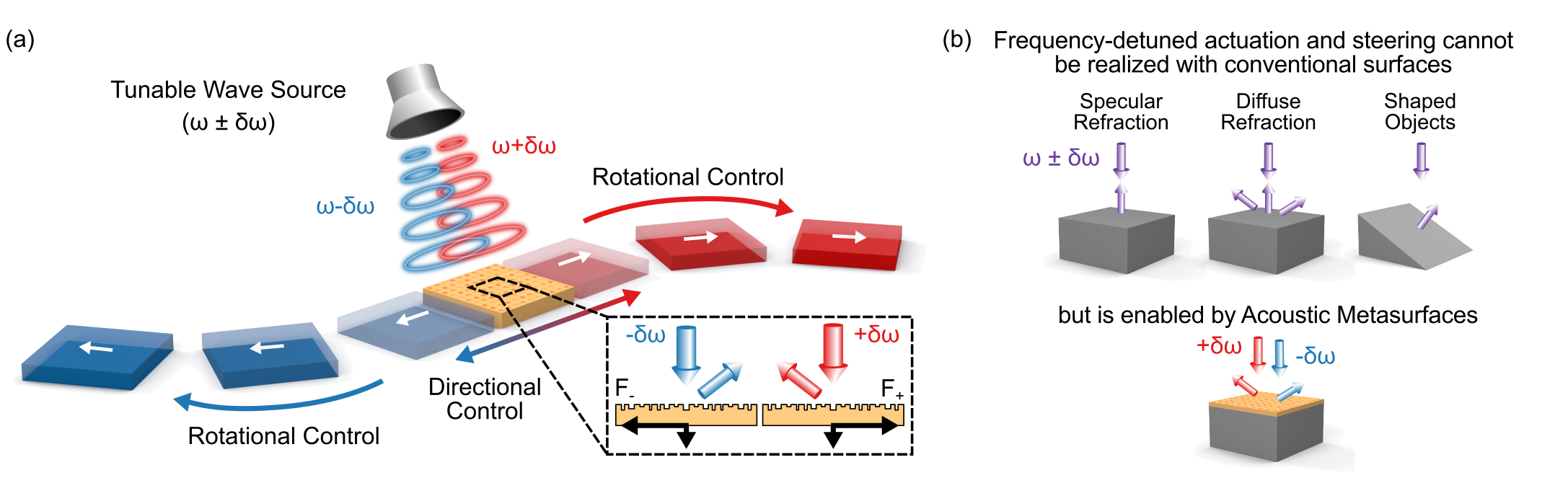}
    \caption{%
        \textbf{Conceptual illustration of programmable control of radiation forces enabled by acoustic frequency detuning:}
        (a) When patterned on the surface of an object, the metasurface converts small and deliberate variations in incident acoustic wave frequency into steerable motion. Frequency detuning ($\pm \delta\omega$, where $\delta\omega/\omega \ll 1$) reverses the sign of the radiation force and steers the motion of a large object ($+\delta \omega$ is red; $-\delta\omega$ is blue). The force/torque is determined by the change in wave momentum, which is controlled by the metasurface local phase. 
        (b) Programmable actuation and steering is not possible with conventional surfaces: these surfaces (whether specular, diffuse, or shaped) exhibit conventional refraction that is fixed by geometry and therefore insensitive to small changes in frequency. Bottom: 
        Metasurface subwavelength structuring imparts frequency-encoded control, enabling contactless and programmable actuation.
    }
    \label{Figure 1}
\end{figure*}
Here, we introduce and demonstrate frequency-tunable metasurfaces that, when patterned on the surface of an object, convert small detunings of the incident wave frequency into steerable, contactless motion of macroscopic bodies that can be much larger than particles.
In this way, force and torque on the object can be directed (and even fully reversed) through remote frequency detuning. 
Metasurfaces---engineered surfaces composed of deep-subwavelength patterns---have emerged as a powerful platform for manipulating wavefronts in beam steering and control
\cite{Xie-Cummer:NatCommun14-o, Ma-Sheng:NatMater14-x, Cummer-Alu:NatureReviews16-u, Wong-Zhang:JOpt17-m, Ge-Sheng:NatlSci18-b, Stein-Ilic:NatCommun22-m}.
Frequency-sensitive metasurfaces have also been explored for multi-frequency applications through frequency multiplexing, where distinct functionalities are assigned to well-separated excitation frequencies~\cite{Assouar-Jing:NatureReviews18-y}.
However, such multiplexed approaches typically require widely spaced frequencies spanning large bandwidths and ratios (e.g., frequency ratios of $\sim 3\times$~\cite{Zhu-Cheng:AIPAdv16-d, Zhang-Ma:MaterHoriz25-l}), limiting their ability to respond continuously to small frequency detunings.
Thus, while metamaterials---both acoustic (\cite{Stein-Ilic:NatCommun22-m, Zhang-Ma:MaterHoriz25-l}) and optical (\cite{Ilic-Atwater:NatPhotonics19-d, Siegel-Brar:ACSPhotonics19-i})---can induce anomalous momentum transport, we demonstrate a mechanism whereby small, deliberate frequency shifts (much smaller than the carrier frequency) produce programmable actuation and steering. 

The fundamental concept is illustrated in Figure \ref{Figure 1}: 
the surface of an object subjected to a wave excitation is decorated with a metasurface that switches the scattered direction under detuned frequencies $\omega\pm\delta\omega$. 
By controlling the direction of scattered waves, we directly control the balance of incident and outgoing momentum and, therefore, the net radiation force vector on the object.
We demonstrate the proposed mechanism theoretically and experimentally, showing fully reversible actuation where the metasurface-decorated object moves in one direction for an incident wave with positively detuned frequency $\omega+\delta\omega$ and in the opposite direction for negatively detuned frequency $\omega-\delta\omega$. 
We report a general design principle for synthesizing such metasurfaces, as well as a strategy to amplify forces by twenty-fold ($\approx 20.9$ times) through systematic optimization of the metasurface topology.
Beyond translation, we extend the concept to rotation, showing that the direction of the torque---and consequently rotation---can also be reversed by frequency detuning.
Our proof-of-concept demonstration focuses on inaudible/ultrasonic frequencies; however, the approach is scalable to other frequencies and wave systems, since the governing phase-geometry relationship depends on the relative size of metasurface features compared to the wavelength, and not on absolute size as seen in Supplementary Figure S1 \cite{supplemental}.
The proposed frequency-controlled actuation mechanism is uniquely enabled by metasurface patterning and cannot be realized with conventional materials (Figure \ref{Figure 1}b).

\section{Results}
To establish the mechanism underlying actuation controlled by frequency-detuning, we model the metasurface as a series of subwavelength unit cells that collectively shape the momentum of the excitation field. 
When unit cells are arranged to impart a phase term $\Phi({\rv})$ to deflect the wave at position $\rv$ on the surface, this anomalous deflection modifies the refracted wave according to the generalized Snell’s law $\sin\theta_R = \sin\theta_I + \frac{1}{k}\frac{\dd\Phi({\rv})}{\dd\rv}$, where $\theta_R$ and $\theta_I$ are the reflected and incident sound wave angle, respectively, and $k$ is the wave-vector. Recasting in momentum form, the generalized Snell’s law determines the lateral force on the metasurface, i.e., $F_x \propto k_x^r - k_x^i$, where $k_x^r$, $k_x^i$ are the wave-vector components of the refracted, incident field and $k_x^r = k_x^i + \partial\Phi/\partial x$ (a similar relationship applies to the y-axis). 
The recipe for constructing a metasurface that encodes frequency-detuning for momentum control can be formulated through a perturbative approach. First, we start with a metasurface comprising two mirror-symmetric halves that, in equilibrium, experiences a balanced (net zero) force at target frequency $\omega$. Next, perturbing the geometric profile of unit cells comprising the left half, we induce the modified phase profile $\Phi_{\textrm{left}}$ to increase the force at the negatively detuned frequency $\bm{F}_{\textrm{left}}(\omega - \delta \omega)$. Similarly, for the right half, we seek to increase the force at positively detuned frequency $\bm{F}_{\textrm{right}}(\omega + \delta \omega)$. To first order, this creates a force imbalance within the $\delta \omega$ frequency bandwidth. Mathematically, we seek the strongest possible opposing forces over the frequency bandwidth, \eg to maximize the product $-\bm{F}(\omega+\delta\omega)\cdot\bm{F}(\omega-\delta\omega)$, where $\bm{F}$ is the force vector on the metasurface (the ''$-$'' sign indicates the forces at $\omega \pm \delta \omega$ are opposing). \\

The key insight is to perturb metasurface topology in a manner that induces (and subsequently enhances) the force-reversing asymmetry.
We propose that the desired asymmetry can be found in the derivative of the spatial phase with respect to the unit cell geometry. Since the force on the metasurface depends on the spatial derivative of the phase $\frac{\dd \Phi(\rv)}{\dd\rv}$, it follows by chain rule that $\frac{\dd \Phi(\rv)}{\dd\rv} = \frac{\dd \Phi(\rv)}{\dd h_j} \cdot \frac{\dd h_j}{\dd\rv}$, where $h_j$ is the geometric parameter of the $j$-th cell that controls the local phase. We keep the spatial variation of $h_j(\rv)$ frequency-independent so that the sign of the lateral force can be frequency controlled through the spatial derivative $\frac{\dd \Phi(\rv)}{\dd h_j}$. Therefore, it becomes possible to assemble a composite metasurface where the left or right half comprises unit cells with a stronger gradient at $\omega+\delta\omega$ ($\omega - \delta\omega$). 

\begin{figure}[t]
    \includegraphics[scale=1]{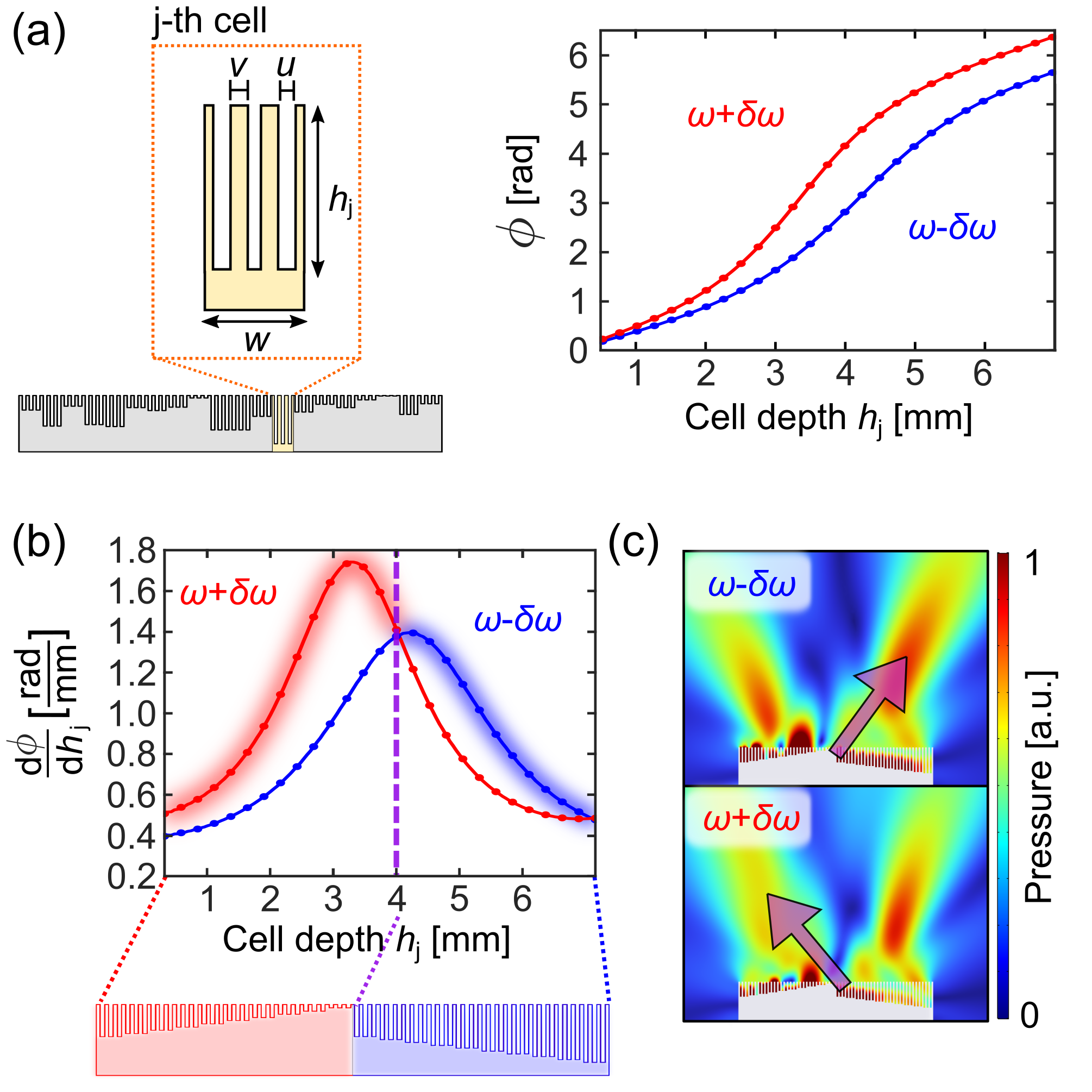}
    \caption{%
        \textbf{Analytic design principle for metasurfaces that encode force control via frequency detuning.}
        (a) Phase response for a candidate unit cell for inaudible frequency $\omega=22.5 \mathrm{kHz}$ and detuning $\delta\omega=2.5 \mathrm{kHz}$ as a function of cell depth $h$. 
        (b) Analytical metasurface constructed from the spatial gradient of phase to achieve directional force control.
        (c) Simulated scattering plot for a downward incident acoustic wave at the two detuned frequencies $\omega \pm \delta\omega$.
    }
    \label{Figure 2}
\end{figure}

Figure \ref{Figure 2} demonstrates this approach for a unit cell with a U-shaped topology. For a building-block cell comprising three grooves, the cell depth $h_j$ controls the local phase. 
This unit cell design serves as a practical starting point because it provides subwavelength phase control while remaining simple to fabricate by additive manufacturing. The three-groove building block is not a fundamental requirement, but a design choice that reduces the number of optimization parameters while preserving sufficient phase tunability to demonstrate frequency-controlled force reversal.
Figure \ref{Figure 2}a shows the reflectance phase profile versus depth for $\omega \pm \delta \omega$, where $\omega=22.5$ kHz and $\delta\omega=2.5$ kHz are chosen to be inaudible/ultrasonic. Unit cell parameters are $w=2.571$ mm, $u = 0.45$ mm, $v=0.407$ mm. While the phase profiles appear similar for nearby frequencies, it is the derivative of phase with respect to cell depth $\frac{\dd \Phi(\rv)}{\dd h_j}$ that shows strong frequency dependence (Figure \ref{Figure 2}b). We observe that cells with smaller depth ($h<4.0$ mm) exhibit a larger gradient at the higher frequency, whereas the opposite is true for deeper cells ($h>4.0$ mm). Therefore, we assemble a composite analytic metasurface where the left (right) side comprises shallower (deeper) cells. Further, we linearly vary cell depth to keep $\frac{\dd h_j}{\dd\rv}$ constant and ensure frequency dependence will be dictated by the phase derivative  $\frac{\dd \Phi(\rv)}{\dd h_j}$. The parameters and dimensions of the analytic metasurface are listed in Supplementary Table S1.

The designed analytic metasurface, as shown in the bottom part of Figure \ref{Figure 2}b, is numerically modeled in COMSOL Multiphysics. The metasurface consists of $N=20$ unit cells, evenly split between the two sides. Figure \ref{Figure 2}c shows the 2D scattering response for the two frequencies. The preferential scattering to the left for $\omega - \delta\omega$ frequency (and vice-versa for $\omega + \delta\omega$ frequency) confirms the designed behavior. To quantify the radiation pressure force on the metasurface, we evaluate the surface integral
$\bm{F} = -\int_S \left[  \left\langle p_2 \right\rangle \bm{n} + \rho_0 \left\langle (\bm{n}\cdot \bm{v}_1)\bm{v}_1 \right\rangle \right]\dd A$, where $p_2$, $\bm{v}_1$ are the first and second order perturbations in pressure and velocity, respectively, $\bm{n}$ is the surface normal unit vector, and $\rho_0$ is the medium mass density
\cite{Glynne-Jones-Hill:JAcoust13-x, Stein-Ilic:NatCommun22-m}. The integral is numerically evaluated in three dimensions over a closed surface surrounding the object (by momentum conservation, the integral can be evaluated over any surface that surrounds the object). 
Of particular interest to us is the ratio of forces at the two frequencies. From the simulation, we find the force ratio to be $F_x (+\delta\omega) \approx -1.52\hspace{2px} F_x(-\delta\omega)$, which confirms the frequency-reversing behavior of the composite analytical metasurface ($F_x$ is the lateral force on the metasurface and $\pm \delta\omega$ is a shorthand for $\omega \pm \delta\omega$). Therefore, this metasurface would move in the positive x-direction under detuned frequency $+\delta\omega$, and similarly in the negative x-direction for $-\delta\omega$ detuning.

Next, we show the force-reversing effect can be greatly enhanced by exploiting the parametric topology of the metasurface. 
We treat the the $N$ unit cells depths $(h_1, h_2, \ \ldots,\ h_N)$ as independent but variable design optimization parameters. 
At each iteration of the optimization process, the force contrast at the two detuned frequencies is evaluated using finite-element analysis in COMSOL Multiphysics. 
Numerically, we seek to maximize the product $-F_x(\mathbf{h};+\delta\omega)\cdot F_x(\mathbf{h};-\delta\omega)$, where $\mathbf{h} = [h_1,\ \ldots,\ h_N]$ (the detuning $\delta\omega = 2.5$ kHz is kept fixed).
Since unit cell depths $\mathbf{h}$ are independent (but fabrication-constrained) design variables, this process produces metasurfaces that become substantially differentiated from the initial analytic description.
However, this leads to a challenging, non-convex design space since neighboring cells are no longer constrained to satisfy a specific phase profile. 
We address this issue with the Subplex optimization algorithm, notable for its efficacy for complex and high-dimensional optimization spaces \cite{Rowan-Rowan:Other90-v}
(the Subplex implementation was integrated through the NLopt package \cite{Johnson_undatedr-s}). 
Each optimization run was initialized from the analytic metasurface shown in Fig. \ref{Figure 2}b, with all $h_j$ values constrained to lie within the manufacturable range $0.5~\mathrm{mm} \le h_j \le 9~\mathrm{mm}$. At every Subplex iteration, COMSOL was used to solve the full-wave acoustic scattering problem and compute the radiation force via the surface integral given above. \\

\begin{figure*}[t]
    \includegraphics[width=\textwidth]{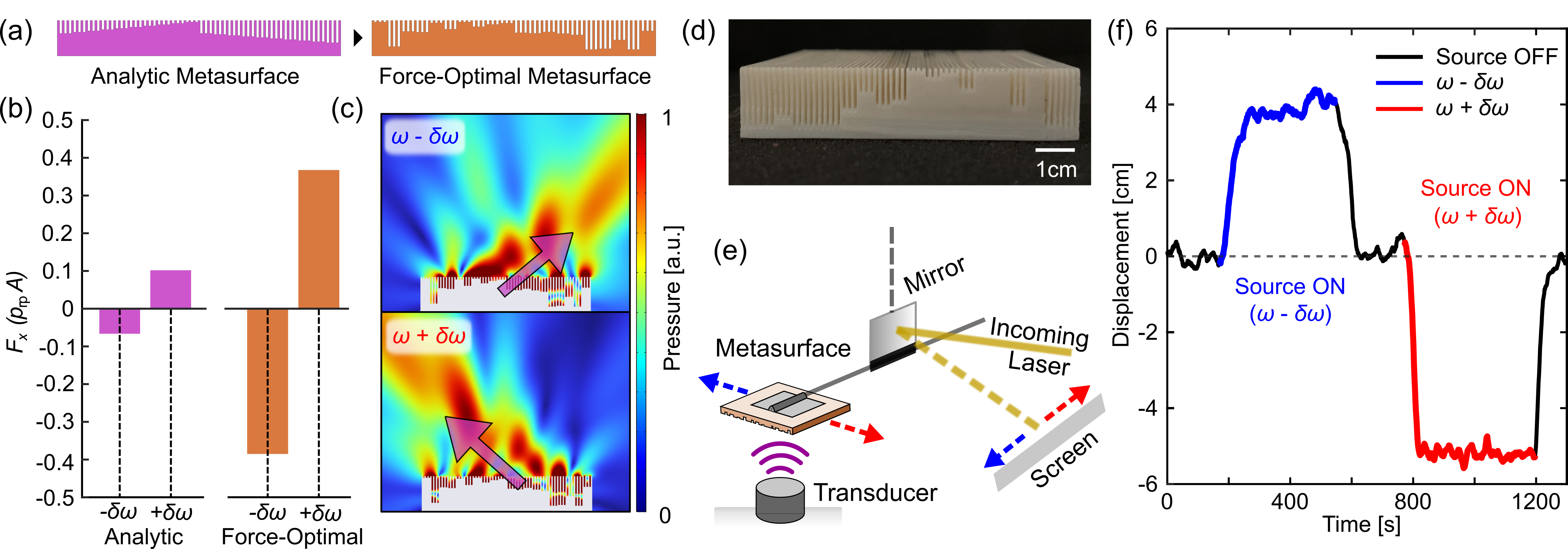}
    \caption{%
        \textbf{Actuation using force-optimized acousto-mechanical metasurfaces: design, fabrication, and experimental results.}
        (a) Analytical and topology-optimized metasurface geometries illustrating how structural asymmetry produces frequency-sensitive forces ($N=20$ unit cells).
        (b) Comparison of force response at the two detuned frequencies ($\pm \delta\omega$). Force-optimal metasurface shows over an order of magnitude enhancement in force contrast, as per Eq. \ref{eq:1}.
        (c) Scattering plot for an incident downward plane wave at the two detuned frequencies ($\omega=22.5 \mathrm{kHz}$, $\delta\omega=2.5 \mathrm{kHz}$).
        (d) Fabricated metasurface prototype.
        (e) Schematic of the experimental setup. The transducer frequency can be controlled and tuned. 
        (f) Experimental time trace of metasurface displacement showing fully reversible motion under detuned excitation. Blue corresponds to transducer emitting at $\omega-\delta\omega$ (displacement is positive); Red corresponds to $\omega+\delta\omega$ (displacement is negative); here, $\omega=17.5 \mathrm{kHz}$, $\delta\omega=2.5 \mathrm{kHz}$. Black denotes that the source is turned off.
    }
    \label{Figure 3}
\end{figure*}

Figure \ref{Figure 3}a shows the profile of the optimized force-reversing metasurface created from this process. 
The geometrical dimensions of the metasurface are listed in Supplementary Table 2 (see Supplemental Material).
The discrete nature of the optimized metasurface arises because the optimization is not constrained to preserve the smooth phase-gradient profile of the analytical design. Instead, the optimizer directly maximizes the radiation-force objective at both $\omega-\delta\omega$ and $\omega+\delta\omega$ using the full finite-element force calculation. This allows the design to exploit discrete scattering and non-intuitive coupling between neighboring unit cells, redirecting more of the incident acoustic momentum into the desired lateral force than the analytical phase-gradient profile.
The overall force is normalized to the cross-sectional area $A$ and the radiation pressure $p_{rp}$ of the wave ($p_{rp}A$ relates to the wave momentum, i.e., the momentum that the wave would transfer to an ideal attenuating surface). 
The scattering response at the two detuned frequencies (Fig. \ref{Figure 3}b) shows an improvement of preferential directional scattering relative to that of the analytic metasurface (Fig. \ref{Figure 2}b).
Most importantly, the force on the metasurface is substantially increased at both frequencies, i.e.,  $F_x^{\textrm{optimized}}(+\delta\omega) \approx 3.6 F_x^{\textrm{analytic}}(+\delta\omega)$ and $F_x^{\textrm{optimized}}(-\delta\omega) \approx 5.8 F_x^{\textrm{analytic}}(-\delta\omega)$, where the radiation force is calculated from the contour integral as before. 
Overall, the improvement in the figure of merit is 
\begin{equation}
    \frac{-\left[F_x(+\delta\omega)F_x(-\delta\omega)\right]^{\mathrm{optimized}}}{-\left[F_x(+\delta\omega)F_x(-\delta\omega)\right]^{\mathrm{analytic}}}
    \approx 20.9
    \label{eq:1}
\end{equation}
This enhancement is obtained by applying the same finite-element force calculation to both the analytical and optimized metasurfaces.
The larger-than-order-of-magnitude enhancement demonstrates the potential of metasurface topology design for realizing target force profiles to achieve programmable and contactless actuation. Although $\delta\omega/\omega\ll1$, the force contrast is amplified by the steep frequency dependence of the optimized metasurface.
To highlight the absolute momentum conversion, we compare the simulated lateral radiation forces to an idealized $45\degree$ reflector (for which the lateral component of the redirected momentum is $70.7\%$ of the incident momentum). Using this value as a reference, the analytical metasurface reaches efficiencies of $9.4\%$ and $14.3\%$ at the two detuned frequencies, whereas the optimized metasurface reaches $54.5\%$ and $52.1\%$. Equivalently, relative to the total incident momentum flux, these correspond to $6.6\%$ and $10.1\%$ for the analytical metasurface and $38.5\%$ and $36.8\%$ for the optimized metasurface (Figure \ref{Figure 3}b). The optimized design reaches more than half of the $45\degree$ lateral-momentum reference at both detuned frequencies, despite relying only on small frequency shifts about the carrier frequency.

We proceed to experimentally demonstrate the force-reversing mechanism. 
For the concept-proof experiment, the center frequency of $\omega=17.5$ kHz and detuning $\delta\omega=2.5$ kHz is selected because tunable acoustic sources, such as high-frequency compression drivers, are readily available in this frequency range. Further, this frequency range is high enough to reduce the sensitivity to the human ear, yet small enough (i.e., large wavelength) to ensure the smallest metasurface features are compatible with the printing resolution limits of conventional 3D printers. 
Figure \ref{Figure 3}d shows the metasurface fabricated using a photopolymer jetting 3D printer. 
The metasurface was printed in VeroWhitePlus photopolymer (density $1.18~\mathrm{g/cm^{-3}}$, modulus of elasticity $\sim 2.5~\mathrm{GPa}$), which acts as a rigid boundary at ultrasonic frequencies.
We followed the same process to redesign the metasurface for the operation at $\omega=17.5$ kHz and detuning $\delta\omega=2.5$ kHz. 
The overall dimensions of the printed metasurface object are $L=51.9$ mm, $D=50.5$ mm, and $H=11.1$ mm, for length, depth, and height, respectively, and the complete metasurface profile is provided in the Supplementary Table 3. 
The experimental setup to measure actuation caused by acoustic waves utilizes the concept of a torsion pendulum (Figure \ref{Figure 3}e). 
When a wave is anomalously refracted by a metasurface, it generates a lateral force, resulting in the metasurface deflection. 
This movement is detected through the displacement of a laser spot on a screen, which is reflected from a mirror at the base of the pendulum (Figure \ref{Figure 3}e). 
A camera captures the screen and feeds images to a spot-tracing algorithm that identifies the location of the laser spot in real time. The data is converted to positional information of the metasurface, thus enabling live tracking of the metasurface actuation (additional details on the experimental method are provided in Supplemental Materials). 

The frequency-controlled force-reversing effect can be seen in Figure \ref{Figure 3}f. In the experiment, we direct the wavefront at the metasurface from a compression driver located approximately 8 cm away. 
Initially, the acoustic source is off, and the metasurface is stationary.
Once the source is switched on and driven at the $\omega-\delta\omega$ frequency, a sharp jump in the metasurface displacement is observed (Fig. \ref{Figure 3}f, solid blue), corresponding to the onset of acoustic radiation pressure. 
As the metasurface begins to deflect, it moves away from the acoustic source, reducing the lateral acoustic force acting on it. The motion continues until the acoustic force is balanced by the restoring torque from the pendulum. Once the sound source is turned off, the acoustic force is removed, and the stored elastic energy in the pendulum returns the metasurface to its original position.
Next, the source is again turned on but driven at the $\omega+\delta\omega$ frequency, this time resulting in a sharp negative displacement (Fig. \ref{Figure 3}f, solid red), corresponding to the metasurface moving in the opposite direction. 
Finally, after turning off the source, the metasurface returns to its original position.
The blue and red colored lines in Figure \ref{Figure 3}f denote the times during which the source is kept on at the $-\delta\omega$ and $+\delta\omega$ detuned frequency, respectively.
Measurements taken with the acoustic source turned off were used to establish the baseline drift and noise level. The optimized metasurface was selected for experimental characterization because its displacement response is well above this baseline. In addition, the repeatable displacement reversal under positive and negative frequency detuning distinguishes the measured response from drift due to ambient air currents or thermal gradients.

The proposed mechanism to control (e.g., reverse and steer) acoustic forces can be directly adapted for contactless actuation at other frequencies.
For example, metasurfaces shown in Figure \ref{Figure 3} were designed to operate at $\omega=22.5$ kHz and $\delta\omega=2.5$ kHz. However, for our experiments, we used the same approach to design a metasurface for operation at $\omega=17.5$ kHz (same $\delta\omega$) to better match the frequency range for which frequency tunable sources are readily available.  
For the latter (experimental) metasurface, Supplementary Figure S1 shows the scattering profile.
The overall force and force contrast at the two detuned frequencies $\omega\pm\delta\omega$, as evaluated in the simulation, shows a comparable magnitude as before.
We selected the U-shaped, three-groove cell for its simplicity of fabrication but other metasurface topologies may offer superior performance, such as stronger force contrast or a thinner/lightweight profile, provided sufficient phase coverage, phase sensitivity and steering efficiency can be achieved.

\begin{figure}[t]
     \includegraphics[scale=0.95]{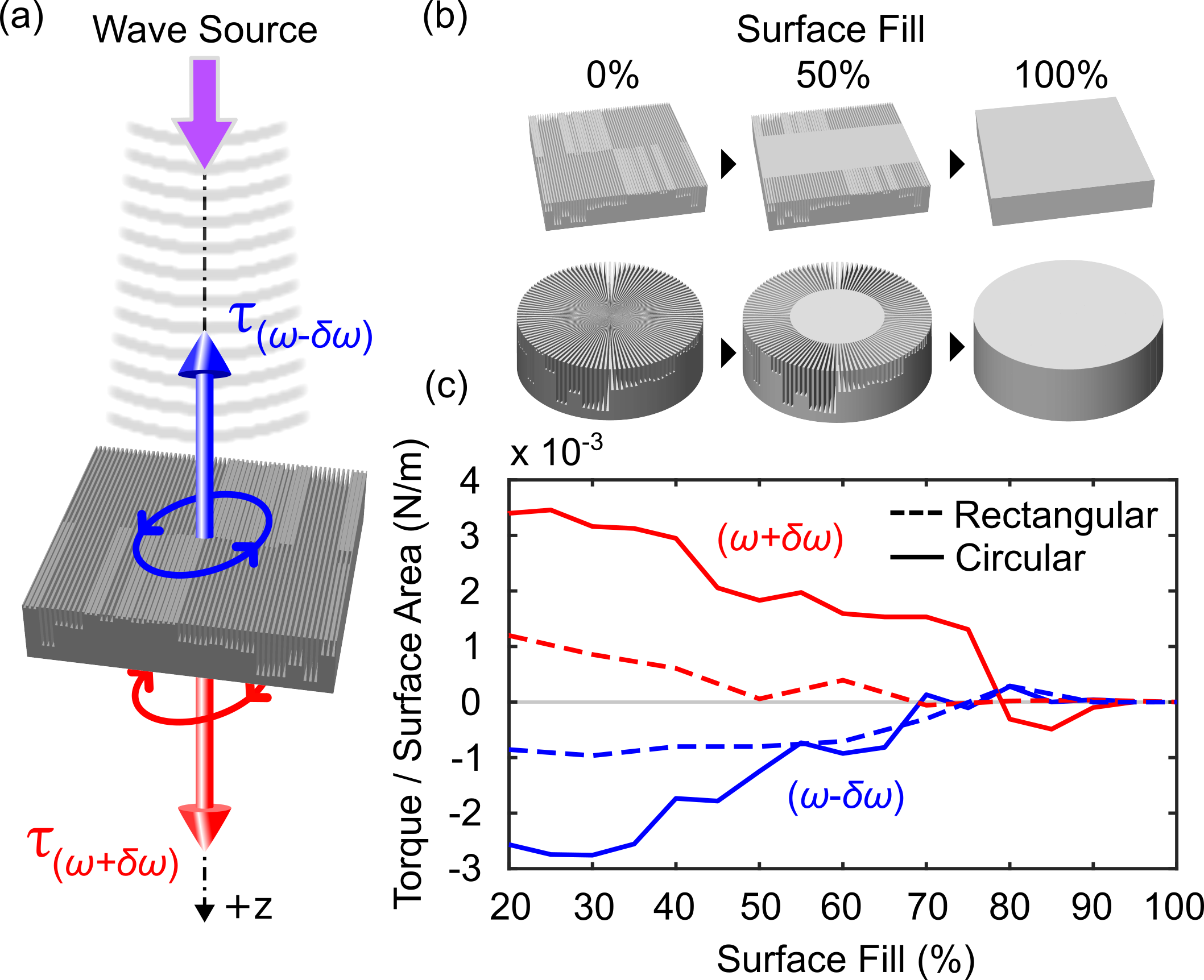}
    \caption{%
        \textbf{Torque control via frequency detuning}.
        (a) Small and deliberate detuning of the acoustic excitation ($\pm\delta\omega$) reverses the induced torque, demonstrating reversible rotational actuation.
        (b) Embedding force-optimized metasurfaces on rectangular and cylindrical objects demonstrates how patterned surface coverage (surface fill) can tailor rotational response while leaving unpatterned regions as space for integrating additional functionality.
        (c) Torque per unit surface area versus surface fill. Blue corresponds to transducer emitting at $\omega-\delta\omega$; red corresponds to $\omega+\delta\omega$, where $\omega=22.5 \mathrm{kHz}$, $\delta\omega=2.5 \mathrm{kHz}$. Solid lines: circular; dashed lines: rectangular configurations.
    }
    \label{Figure 4}
\end{figure}

Next, extending the principle of frequency-encoded actuation, we show that acousto-mechanical metasurfaces can reversibly control torque, establishing a route toward programmable rotational control.
 This concept is highlighted in Figure \ref{Figure 4}a, where the object experiences a clockwise torque under $-\delta\omega$ frequency detuning and a counter-clockwise torque for $+\delta\omega$ detuning. 
The force-optimal metasurface topology presented in Figure \ref{Figure 3}a is used to pattern both rectangular and circular objects, as seen in Figure \ref{Figure 4}b, and torque as a function of surface area is evaluated using the same closed-contour integral in acoustomechanical simulations. 
The rectangular object is patterned with two force-optimal metasurfaces, each oriented to generate lateral forces in opposing directions under the two detuned frequencies. Due to the axial symmetry of the object, these opposing forces result in torque and rotation control. 
The circular object is patterned by wrapping the force-optimal metasurface topology around its circumferences. 
Because the U-shaped unit cells are intrinsically three-dimensional, the circular design is radially scaled to conform to the object curvature and enhance torque per unit area. 

To illustrate the multifunctional potential of torque metasurfaces, we incorporated solid, unpatterned regions within both rectangular and circular designs. These unpatterned domains offer space for embedding auxiliary functionalities---such as sensing, control, or additional actuation---while preserving rotational steering. 
We further explore the balance between reserving area for such multifunctionality and the resulting reduction in torque.
This concept is visualized in Figure \ref{Figure 4}b, where surface fill is initially constrained to the center of the object and increased to encompass the entire active surface.
It was observed that as surface fill increases, the total torque per unit surface area decreases until the entire surface becomes void of any patterning, resulting in zero net torque at either frequency.
Additionally, both circular and rectangular objects experience near-constant torque up to 30\% surface fill, as observed in Figure \ref{Figure 4}c. 
At low surface fill percentages, we observe a nearly three-fold increase in torque per surface area when comparing circular to rectangular metasurfaces. 
We attribute this to the more efficient conversion of in-plane force into axial torque in a radial configuration relative to a rectangular configuration.
Generally, the maximum load that can be rotated will depend on the acoustic source intensity, object size, and the surrounding medium. In the present design, the achievable torque is constrained by the lateral radiation force generated by the patterned surface and by the moment arm over which that force acts. The rotation results should therefore be interpreted as a functional demonstration of torque reversal using the force-optimized design from Fig. \ref{Figure 3}a, rather than as a globally optimized design for rotational actuation. Further improvements could be achieved by directly optimizing the metasurface unit-cell pattern for torque.
\section{Conclusions}
In conclusion, this work demonstrated that acoustic waves can induce controllable, and steerable, translation and rotation of macroscopic objects via small, but deliberate, detuning of wave frequency. 
Uniquely enabled by metasurface wavefront manipulation, this approach allows for dynamical behaviors to be first designed by simply altering the patterns on the surface and then controlled by detuning of the wave field frequency.
While this work primarily focused on translation and rotation along a single axis, the concept could be extended to two-dimensional metasurfaces for actuation along several axes (e.g., multi-axial steering) by encoding $\partial\Phi/\partial x$ and $\partial\Phi/\partial y$ within the same surface. 
Such a design would require unit cells with additional geometric degrees of freedom (e.g., independently tunable depth and width) to encode frequency-dependent phase gradients along both lateral axes, and the optimization framework employed here can be extended to this higher-dimensional design space. The overall metasurface footprint would accordingly need to be comparable in both lateral dimensions, in contrast to the elongated geometry of the present one-dimensional design.
In addition to rigid thermoplastics used for metasurface fabrication in this work, exploring the use of flexible materials could enable novel ways of controlling forces in soft material systems and soft robotics \cite{Bertoldi-vanHecke:NatureReviews17-z, Jin-Brunet:NatCommun19-j, Nassar-Haberman:NatureReviews20-h}. 
The frequency-detuning principle could also be extended to liquid or tissue-like environments, with the same objective of using small frequency detunings to produce large, reversible changes in acoustic radiation force. Such implementations would require medium-specific redesign, since impedance contrast, viscous damping, and material losses would modify the unit-cell phase response and momentum-transfer efficiency. Similar constraints have already been addressed in acoustic metasurfaces for underwater operation, suggesting a path toward adapting this approach beyond air.
\\

A particularly intriguing question pertains to the size of the metasurface, specifically, the number and the complexity of cells needed to realize the desired frequency-controlled force profile. 
From this work, it is clear that metasurfaces with many cells (e.g., consisting of tens of cells) are adequate, but how the force profile varies with overall metasurface size remains to be systematically explored.
We hypothesize that analytic designs may likely necessitate more cells, but topology-optimized surfaces could potentially be much more efficient in a smaller size.
Looking forward, the ability to realize programmable forces and torques in constrained or miniaturized geometries---as well as in flexible and lightweight substrates---could open impactful opportunities for battery-free (micro)robotic devices, biomedical interfaces, and reconfigurable soft matter systems. 
By bridging wave control with mechanical actuation, these acousto-mechanical metasurfaces establish an approach for multifunctional, adaptive, and programmable actuation across scales and material platforms.

\section{Acknowledgements}
\vskip 2ex
\begin{acknowledgments}
    This material is based upon work supported by the National Science Foundation (NSF) under Grant No. CMMI-2318094.
    The authors also acknowledge support from the Air Force Office of Scientific Research (AFOSR) under Grant No. FA9550-22-1-0070.
\end{acknowledgments}

\textbf{Data Availability Statement}: The data supporting the findings of this study are available within the main article and its supplementary materials. Raw data are available on Zenodo at \url{https://doi.org/10.5281/zenodo.20619402}.

*These authors contributed equally to this work.

$^{\dag}$ E-mail: ilic@umn.edu


\FloatBarrier
\bibliographystyle{apsrev4-2-longbib}

\end{document}